\documentclass{elsarticle}

\usepackage[utf8]{inputenc}
\usepackage[russian,english]{babel}

\usepackage{amsmath}
\usepackage{bm}
\usepackage{tikz}
\usepackage{pgfplots}

\usepackage[pdfborder={0 0 0}]{hyperref}

\newproof{pf}{Proof}

\newcommand{\grad}{\mathop{\rm grad}\nolimits}
\renewcommand{\div}{\mathop{\rm div}\nolimits}

\newcommand{\tensor}[1]{\bm{#1}}
\renewcommand{\vec}[1]{\bm{#1}}

\journal{arXiv.org}

\begin{document}

\begin{frontmatter}

\title{Numerical simulation of the stress–strain state of the
dental system}

\author[im]{Sergey~V.~Lemeshevsky}
\ead{sergey.lemeshevsky@gmail.com}

\author[bu]{Semion~A.~Naumovich}
\ead{ortopedstom@bsmu.by}

\author[bu]{Sergey~S.~Naumovich}
\ead{serg.nm@mail.ru}

\author[nsi,univ]{Petr~N.~Vabishchevich\corref{cor}}
\ead{vabishchevich@gmail.com}

\author[univ]{Petr~E.~Zakharov}
\ead{zapetch@gmail.com}

\address[im]{Institute of Mathematics, National Academy of Sciences, Minsk, Belarus}
\address[bu]{Belarusian State Medical University, Minsk, Belarus}
\address[nsi]{Nuclear Safety Institute, Russian Academy of Sciences, Moscow, Russia}
\address[univ]{North-Eastern Federal University,  Yakutsk, Russia}

\cortext[cor]{Corresponding author}

\begin{abstract}
We present mathematical models, computational algorithms and software,
which can be used for prediction of results of prosthetic
treatment. More interest issue is biomechanics of the periodontal
complex because any prosthesis is accompanied by a risk of overloading
the supporting elements. Such risk can be avoided by the proper load
distribution and prediction of stresses that occur during the use of
dentures. We developed the mathematical model of the periodontal
complex and its software implementation. This model is based on linear
elasticity theory and allows to calculate the stress and
strain fields in periodontal ligament and jawbone.

The input parameters for the developed model can be divided into two
groups. The first group of parameters describes the mechanical
properties of periodontal ligament, teeth and jawbone (for example, elasticity of
periodontal ligament etc.). The second group characterized the
geometric properties of objects: the size of the teeth, their spatial
coordinates, the size of periodontal ligament etc.
The mechanical properties are the same for almost all, but the input
of geometrical data is complicated because of their individual
characteristics. In this connection, we develop algorithms and
software for processing of images obtained by computed tomography (CT) scanner
and for constructing individual digital model of the
tooth-periodontal ligament-jawbone system of the patient.
Integration of models and algorithms described allows to carry out
biomechanical analysis on three-dimensional digital model and to
select prosthesis design.
\end{abstract}


\end{frontmatter}

\section{Introduction}

Nowadays, theoretical studies of applied problems are performed on the
basis of the extensive use of computational tools (computers and
numerical methods) \cite{Sam2001,Hri2003}.  Here, the up-to-date
concept of the so-called component-based software
\cite{wang2005component,liu2006mathematical} is discussed.
Component-based software is a set of well-developed software
components, which solve individual basic problems.  Computational
technologies for scientific researches are based on constructing
geometrical models, generating computational meshes, applying
discretization methods, approximate solving of discrete problems,
visualizing, and processing calculated data.

Information technology are used in all areas of medicine, and
dentistry is not exception \cite{book:1078699}. Unfortunately, often
most doctors use computers only to maintain office and medical
documentation, and take it as the information technology.  At the same
time, information technologies are widely used in dentistry for
diagnostics: digital X-rays, digital photos, scanned diagnostic
model. The production of dental prostheses based on computer milling
\cite{book:1375379} as well as various system in the gnathology are
developing very rapidly.

In our view, it is very promising the development of software that
combine processing various images of dental system and subsequent
treatment planning.
To justify the choice of designs of dental prostheses and orthodontic
appliances the methods of mathematical modelling can be used. The
essence of this methods is the ability to predict and evaluate the effect of
medical intervention using the calculation of stress-strain state
of prostheses and appliances as well as tissues and organs of dental
system. Unfortunately, the use of mathematical modelling methods is
usually limited to research with practical recommendations. In our
view, this approach has right to life. However, the rapid development of
information technology and the various methods of diagnosis allows us
to bring the mathematical modelling to the doctor.

The use of mathematical modelling in practical work of dentists is
limited to a number of reasons, the main of which is the lack of
appropriate software. Most of the mathematical models described in the
literature are implemented using applied software packages for finite element
analysis. Operation with them requires special skills and knowledge as
well as involvement of specialists in mathematics. Moreover, these
software packages are aimed at solving a wide range of problems, which
greatly increases their price.

In this paper we present the software package that allows to carry out
the biomechanical analysis of dental system. The basis of the software
is the mathematical model that allows to calculate the stresses and
displacements in teeth, periodontal ligament and jawbone for the given geometric
models and loads on the prosthesis. The software provides an
opportunity of constructing the geometrical model of the dental system
based on processing CT images.

\section{Applied software}

Traditionally we recognize two types of software: applied software and system software.
The latter is supporting software for developing general-purpose applications, 
which is not directly related to applied problems.
Below we consider problems associated with software for numerical analysis of applied mathematical models.
We discuss both commercial and free/open source software for multiphysical simulations.

\subsection{Features of applied software} 

At early stages of using computers, mathematical models were enough simple and insufficiently reliable. 
Applied software, in fact, were primitive, too.
The transition to a new problem or new version of calculations required assembling practically a new program,
which is an essential modification of the previously developed codes.

The modern state of using computational tools is characterized by
studying complex mathematical models.  Therefore, software expands
greatly, becomes large as well as difficult in study and use.
Contemporary software includes a large set of various program units,
which require to construct a certain workflow for efficient
employment.

In numerical simulations, we do not study a particular mathematical problem, but we investigate a class of problems, i.e., 
we highlight some order of problems and hierarchy of models.
Therefore, software should be focused on multitasking approach and be able to solve a class of problems via quick
switching from one problem to another. 

Numerical analysis of applied problems is based on multiparametric predictions.
In the framework of some specific mathematical model, it is necessary to trace the impact of various parameters 
on the solution.
This feature of numerical simulations requires that software have been adapted to massive calculations.

Thus, software developed for computational experiments must, on the one hand, be adapted to quick significant modifications,
and, on the other hand, be sufficiently conservative to focus on massive calculations using a single program.

Our experience shows that applied software is largely duplicated during its lifecycle.
To repeat programming a code for a similar problem, we waste time and eventually money.
This problem is resolved by the unification of applied software and its standardization both globally 
(across the global scientific community) and locally (within a single research team).
 
\subsection{Modular structure}

A modular structure of applied software is primarily based on a modular analysis of the area of applications. 
Within a class of problems, we extract relatively independent sub-problems,
which form the basis to cover this class of problems, i.e., each problem of the class can generally be treated as 
a certain construction designed using individual sub-problems.

An entire program is conceptually represented as a set of modules appropriately connected with each other.
These modules are relatively independent and can be developed (coded and verified) separately.
The decomposition of the program into a series of program modules implements the idea of structured programming.

A modular structure of programs may be treated as an information graph, which vertices are identified with program modules,
and branches (edges) correspond to the interface between modules.

The functional independence and content-richness are the main requirements to program modules.
A software module can be associated with an application area.
For example, in a computational program, a separate module can solve some meaningful sub-problem.
Such a module can be named a subject-oriented module.

A software module can be connected with the implementation of some computational method,
and therefore the module can be called a mathematical (algorithmic) module (solver).
In fact, it means that we conduct a modular analysis (decomposition) at the level of applications or
computational algorithms.

Moreover, mathematical modeling is based on studying a class of mathematical models and, in this sense,
there is no need to extract a large number of subject-oriented modules.
A modular analysis of a class of applied problems is carried out with the purpose to identify 
functionally independent individual mathematical modules.
   
Software modules may be parts of a program, which are not directly related to some meaningful
in applied or mathematical sense sub-problem. They can perform some supporting operations and functions.
This type of modules, referred to as internal ones, includes data modules, documentation modules, etc.

Separate parts of the program are extracted for the purpose of autonomous (by different developers)
designing, debugging, compiling, storage, etc.
A module must be independent (relatively) and easily replaced. 
When we create an  applied software for computational experiments,
we can highlight such a characteristics of software module as actuality, i.e.,
its use as much as possible in a wider range of problems of this class.

\subsection{Basic components}

The structure of applied software is defined according to solving problems.
For modeling multiphysics processes, both general-purpose and specialized software packages are used.

The basic components of software tools for mathematical modeling are the following:
\begin{description}
 \item[\textbullet]  pre-processor --- preparation and visualization of input data (geometry, material properties),
  assembly of computational modules,
 \item[\textbullet]  processor --- generation of computational grids, numerical solving discrete problems,
 \item[\textbullet] post-processor --- data processing, visualization of results, preparation of reports.
\end{description} 
Below we present an analysis of these components. 

\subsection{Data preparing}

In research applied software, the data input is usually performed manually via editing text input files. 
A more promising technology is implemented in commercial programs for mathematical modeling. 
It is associated with the use of a graphical user interface (GUI).

To solve geometrically complex problems, it is necessary to carry out an additional control for input data. 
According to modular structure of applied software, the problem of controlling geometric data is resolved 
on the basis of unified visualization tools.

The core of a pre-processor is a task manager, which provides assembling an executable program for a particular problem.
It is designed for automatic preparation of numerical schemes for solving specific problems.
The task manager includes system tools for solving both steady and time-dependent problems
as well as adjoint problems using a multiple block of computational modules.
A numerical scheme for adjoint problems includes an appropriately arranged chain of separate computational modules.

\subsection{Computational modules}

Software packages for applied mathematical modeling consist of a set of computational modules,
which are designed to solve specific applied problems, i.e., to carry out the processor functionality
(generating mesh, solving systems of equations). 
These software modules are developed by different research groups with their own traditions and programming techniques.

A program package for applied mathematical modeling is a tool for integrating the developed software
in a given application area. The possibility of integrating a computational module into a software system
for mathematical modeling, which makes possible automated assembling computational schemes from separate 
computational modules, is provided by using unified standards of input/output.

A computational module is designed to solve a particular applied problem.
For transient problems, a computational module involves solving the problem from an initial condition up to the final state.
The reusability of computational modules in different problems results from the fact that the computational module
provides a parametric study of an individual problem. 
To solve the problem, we choose a group of parameters (geometry, material properties, computational parameters, etc.),
which can be variated.

System tools of the software system provide a user-friendly interface to computational modules on the basis 
of dialog tools for setting parameters of a problem.

\subsection{Data processing and visualization}

A post-processor of a software system serves for visualization of computational data. 
This problem is resolved by using a unified standard for the output of computational modules.

Software systems for engineering and scientific calculations require
visualization of \texttt{1D}, \texttt{2D}, and \texttt{3D} calculated
data for scalar and vector fields.

Data processing (e.g., evaluation of integral field values or critical parameters) 
is performed in separate computational modules. 
Software systems for mathematical modeling make possible to calculate these additional data
and to output the results of calculations as well as to include them in other documents and reports.

\subsection{User-friendly interface}

In modern applied software, the user interaction with a computer is
based on a graphical user interface (GUI) is a system of tools for
user interaction with a computer, which it based on the representation
of all available user system objects and functions in the form of
graphic display components (windows, icons, menus, buttons, lists,
etc.).  The user has the random access (using keyboard or mouse) to
all visible display objects.

GUI is employed in commercial software for mathematical modeling, 
such as \textsf{ANSYS}\footnote{\url{http://www.ansys.com}}, \textsf{Marc}, \textsf{SimXpert} and 
other products of MSC Software\footnote{\url{http://www.mscsoftware.com}}, 
\textsf{STAR-CCM+} and \textsf{STAR-CD}
from CD-adapco\footnote{\url{http://www.cd-adapco.com}}, 
\textsf{COMSOL Multiphysics}\footnote{\url{http://www.comsol.com}}.
In this case, preparing geometrical models, selecting equations, setting boundary and initial conditions is conducted in
the most easy and clear way for inexperienced users.

\subsection{Componet-based implementation of functionalities} 

Functionality of modern software systems for applied mathematical modeling must reflect the achieved level 
of the development of the theory and practice for numerical algorithms and software.
This goal is achieved by component-oriented programming, which is based on using well developed and 
verified software for solving basic mathematical problems (general mathematical functionality). 

In applied software packages, the actual content of computational problems is the solution of initial value problem 
(the Cauchy problem) for systems of ordinary differential equations, which reflects mass, momentum, and 
energy conversation laws. 
We come to Cauchy problems by discretization in space (using finite element methods, 
control volume or finite difference schemes). 
The general features of the ODE system are the following:
\begin{itemize}
 \item[\textbullet]	it is nonlinear,  
 \item[\textbullet] 	it is coupled  (some valuables are dependent on other ones),
 \item[\textbullet]  it is stiff, i.e., different scales (in time) are typical for the considering physical processes.  
\end{itemize} 
~~~~
To solve the Cauchy problem, we need to apply partially implicit schemes (to overcome stiffness) 
with iterative implementation (to avoid nonlinearity) as well as partial elimination of variables 
(to decouple unknowns). 

Applied software allows to tune iterative processes in order to  take into account specific features of problems: 
to arrange nested iterative procedures, to define specific stopping criteria and  time step control procedures and so on.
In this case, computational procedures demonstrate a very special character and work in a specific 
(and often very narrow) class of problems; the transition to other seemingly similar problems can decrease its efficiency
drastically.

\subsection{Computational components}

Obviously, it is necessary to apply more efficient computational algorithms in order to implement 
complicated nonlinear multidimensional transient applied models taking into account parallel architecture 
(cluster, multicore etc.) of modern computing systems.
The nature of used algorithms must be purely mathematical, i.e., without involving any other considerations.

Here is the hierarchy of numerical algorithms complexity increasing top down:
\begin{description}
 \item[\textbullet] linear solvers -- direct methods and, first of all, iterative methods if systems of equations have large dimensions,  
 \item[\textbullet] nonlinear solvers -- general methods for solving nonlinear systems of algebraic equations,  
 \item[\textbullet] ODE solvers -- for solving stiff systems of ODEs appearing in some mathematical models.
\end{description} 

Solvers must support both sequential and parallel implementations and also reflect modern achievements of 
numerical analysis and programming techniques.
It means, in particular, that we need to use appropriate specialized software developed by specialists in 
numerical analysis.
This software must be deeply verified in practices and greatly appreciated by international scientific community.

Such applied software systems are presented, in particular, in the following collections:
\begin{itemize}
 \item[\textbullet]	\textsf{Trilinos}\footnote{\url{http://trilinos.sandia.gov}} -- Sandia National Laboratory,
 \item[\textbullet] \textsf{SUNDIALS}\footnote{\url{https://computation.llnl.gov/casc/sundials/main.html}} -- 
 Lawrence Livermore National Laboratory,
 \item[\textbullet]  \textsf{PETSc}\footnote{\url{http://www.mcs.anl.gov/petsc}} --  Argonne National Laboratory.
\end{itemize}

\section{Construction of digital geometric model}

To carry out the mathematical modelling stress--strain state of dental
system we have to construct digital geometric model containing all the
computational domains.

\subsection{Segmentation of CT images}

The input data for construction of the digital geometric model is CT
images. Using this data we need to obtain the computational domain
containing sub-domains described below
(Section~\ref{sec:2}). Processing CT images of dental system and
constructing digital geometric models of teeth and jawbone is quite
challenging. This is due to similar X-ray properties of teeth and jawbone,
as well as low (for identification of periodontal ligament) resolution of
contemporary CT scanners. The process of selecting objects in image is
called \emph{segmentation}. We perform segmentation as follows:
\begin{enumerate}
\item Selection of the region of interest.
\item Separation of entire dentition (as a single object \emph{Dentition})
  and upper or lower jawbone (as object \emph{Jaw})
\item Partitioning the object \emph{Dentition} into series of
  individual objects \emph{Tooth\_XX}.
\end{enumerate}

In the first step dentist selects a region of interest on the
loaded CT image.

The second step is constructing models of the dentition and jawbone
using threshold transform and watershed transform with markers. Let us
give a brief description of the used segmentation algorithms.

\subsection{Computational segmentation algorithm} 
\begin{figure}[h]
  \centering
  \includegraphics[width=.75\linewidth]{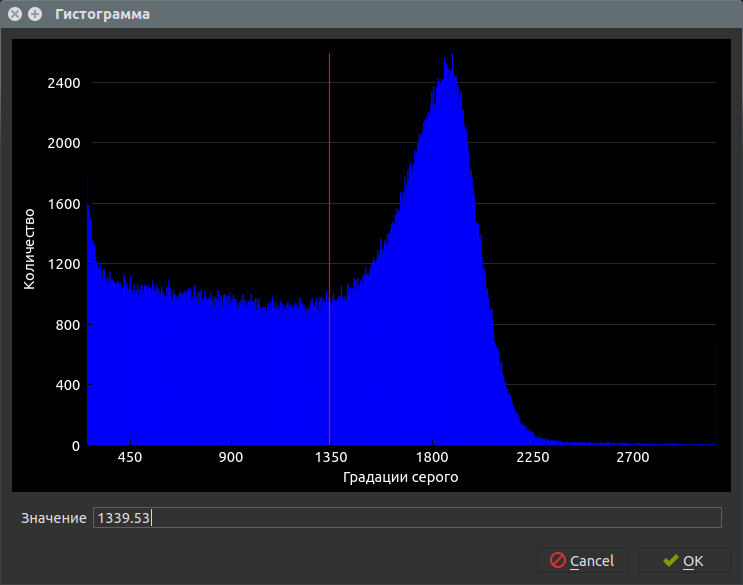}
  \caption{Image histogram}
  \label{fig:1}
\end{figure}

Threshold transform is the basis in applied problems of
segmentation. Suppose that histogram shown in Figure~\ref{fig:1}
corresponds to some image $f(x, y, z)$ containing bright objects on a
dark background, so that the brightness values of the object and the
background pixels are concentrated near some prevailing value. The
obvious way to select an object from the surrounding background is to
choose the threshold level delineating the distribution of brightness
values. In our software user can set the threshold value $T$ (as shown
in Figure~\ref{fig:1}). 

\begin{figure}[h]
  \centering
  \includegraphics[width=.75\linewidth]{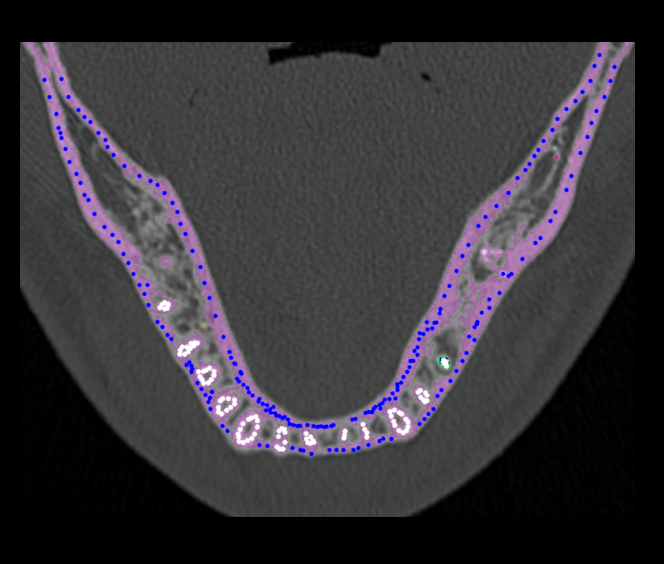}
  \caption{Result of the threshold transform and setting markers}
  \label{fig:2}
\end{figure}

In this case, the result of the threshold transform will contain only
the most optically dense part of pixels in the image of dental system:
teeth and dense jawbone (Figure~\ref{fig:2}, the area colored pink).

Further, to select dentition from jawbone we use approach based on
so-called \emph{morphological watershed transform with markers} that
often give more stable results of the image segmentation. Here we give
only description of the approach. Justification and detailed
exposition of the algorithm can be found in
\cite{GonsWoods,BeareLehmann2006}.

The concept of watershed is based on representation of the image as
three-dimensional surface, given by two spatial variables and
brightness level as height of the surface (terrain). In such a
``topographic'' interpretation we consider the points of three types:
(a) local minimum points; (b) point on the slopes, from which the
water slides in the same local minimum; and (c) points on the ridge,
from which the water slides in more that one local minimum with equal
probability. With regard to specific local minimum the set of points
satisfying condition (b) is called \emph{catchment bassin} of this
minimum. The set of points satisfying condition (c) form ridge lines
on the surface of terrain and are called \emph{watershed lines}.

The main purpose of the segmentation algorithms based on concepts
introduced is to find watershed lines. The basic idea of the method
seems simple. Suppose that hole is pierced in every local
minimum. Then each local minimum becomes the source of a lake. The
entire terrain is progressively flooded through holes and lakes
eventually meet neighbouring lakes. Virtual dams are constructed to
keep the neighboring lakes as the water level rises. When the image
surface is completely flooded the virtual dams correspond to the
watershed lines.

Direct application of the watershed transform usually leads to
excessive segmentation caused by noise and other local irregularities
in the image. Wherein excessive segmentation becomes so significant
that makes the result practically useless. In our case, this means a
large number of domains identified during segmentation. The practical
solution of this problem is to limit allowable number of domains by
including in the procedure preprocessing step, which serves to
introducing additional knowledge.

Approach used to control excessive segmentation is based on the use of
\emph{markers}. Marker is a connected component belonging the
image. We distinguish internal markers, belonging to the object of
interest, and external markers, related to the background. In our
case, the input image is the image obtained by threshold transform and
containing jawboneand dentition only. Therefore, the object \emph{Dentiotion}
corresponds to the internal markers and the object \emph{Jaw} matches the
external markers. At that, setting of markers are carried out by
user. Figure \ref{fig:2} illustrates set of markers on one slice of CT
image.

\subsection{Software implementation} 

For software implementation of the segmentation based on watershed
transform with markers we use cross-platform open source software
\textsf{Insight Segmentation and Registration Toolkit} (\textsf{ITK})
\footnote{\url{http://www.itk.org}} that provides users and developers by
the tools of image processing \cite{ITKSoftwareGuideSecondEdition}.

The resulting rough model of dentition (object \emph{Dentiotion}), which
is obtained by watershed transform, requires clarification. Since the
teeth are normally in contact on the proximal surfaces, then the
object \emph{Dentiotion} is a single object without division into
individual teeth. Therefore the next step is cutting the object
\emph{Dentiotion}. This step is implemented as a special tool, by which
the user cuts the object \emph{Dentiotion} and assigns a label to each tooth.

As a result of segmentation it is obtained the image containing object
\emph{Jaw} and series of objects \emph{Tooth\_XX} (\emph{XX} is corresponding
number of the tooth). This image is saved in \textsf{NIfTI} format
\footnote{\url{http://niftilib.sourceforge.net}}, which is developed by
NIfTI Data Format Working Group and being the adaptation of the
well-known \textsf{Analyze 7.5} format.

\section{Mathematical model of the stress--strain state of dental
  system}
\label{sec:2}

The problem of finding the strain field in dental system is
static elasticity problem. Governing equations of the elasticity are
the momentum balance equation and constitutive equations of elastic medium
\cite{LandauLifshic1986,ciarlet1993mathematical}.

\subsection{Governing equation}

In the case of small strains the momentum balance equation has the
following form:
\begin{equation}
  \label{eq:35}
  \rho \frac{\partial^2 \vec{u}}{\partial t^2} = \div\tensor{\sigma},
\end{equation}
where $t$ is time, $\vec{u}$ is displacement vector,  $\rho$~is
density, $\tensor{\sigma}$~is stress tensor.

The left-hand side of equation \ref{eq:35} (which is inertial term)
becomes significant only in the case of acoustic phenomena, i.e. at
times specific to the period of acoustic properties of the
oscillations. Therefore, in the case of slowly varying loads, inertial
term can be omitted and equation (\ref{eq:35}) takes the form
\begin{equation}
  \label{eq:36}
  \div \tensor{\sigma} = 0.
\end{equation}
The constitutive equations of elastic medium define the relation
between stress and strain ($\tensor{\varepsilon}$) tensors as follows
\begin{equation}
  \label{eq:37}
  \tensor{\varepsilon} = \frac{1}{2} \left( \grad \vec{u} + \left( \grad
      \vec{u} \right)^{\mathtt{T}} \right),
\end{equation}
where $\vec{u}$ is the strain vector field determined for all points
of the medium, subscript $\mathtt{T}$ denotes the transpose of the tensor.

Hooke's law in three-dimensional case can be written as follows
\begin{equation}
  \label{eq:38}
  \tensor{\sigma} = 2\mu(\vec{x}) \tensor{\varepsilon} +
  \lambda(\vec{x}) \div \vec{u} \mathbf{I},
\end{equation}
where $\lambda(\vec{x})$ and $\mu(\vec{x})$ are the Lam\`e coefficients,
$\mathbf{I}$ is the second-rank identity tensor.

Modeled object consists of jawbone ($\Omega_1$), periodontal ligament
($\Omega_2$), teeth, which is supports for bridge prosthesis
($\Omega_3$), and bridge prosthesis ($\Omega_4$). Therefore Lam\`e
coefficients are piecewise constant:
\begin{equation}
  \label{eq:2.5}
  \lambda(\vec{x}) = \lambda_\alpha, \quad \vec{x} \in \Omega_\alpha,
\end{equation}
\begin{equation}
  \label{eq:2.5}
  \mu(\vec{x}) = \mu_\alpha, \quad \vec{x} \in \Omega_\alpha.
\end{equation}

Equations (\ref{eq:36})--(\ref{eq:38}) are the equations of the
elasticity theory, which is supplemented by the boundary conditions on
the outer surface of the modeled object.

\subsection{Boundary conditions} 

To compute stresses arising in dental system under loads on prostheses
we do not need consider whole jawbone, it is known that elastic
stresses, arising from the local loads, decay at distance about the
size of the area of load application. Thereby we may consider only
part of the jawbone extending for distance about $1.5$--$2$ length of
the tooth root from the prosthesis supporting teeth. Therefore we cut the
jawbone by some planes. Thus we get the jaw fragment, the size of
which is determined according to the above criteria.

Then outer surface $\Gamma$ of the jaw fragment separate into three
regions with different external influences:
\begin{enumerate}
\item free surface where external stresses are absent:
  \begin{equation}
    \label{eq:2.1}
    (\tensor{\sigma}\cdot \vec{n})(\vec{x}) = 0, \quad \vec{x} \in \Gamma_1 ;
  \end{equation}
\item surface where strains are absent: 
  \begin{equation}
    \label{eq:2.2}
    \vec{u}(\vec{x}) = 0, \quad \vec{x} \in \Gamma_2 ;
  \end{equation}
\item surface where distributed external load are geven:
  \begin{equation}
    \label{eq:2.3}
    (\tensor{\sigma}\cdot \vec{n})(\vec{x}) = \vec{f}(\vec{x}), \quad \mathbf{x} \in \Gamma_3 .
  \end{equation}
\end{enumerate}

\subsection{Finite element discretization}

Let us perform discretization of problem (\ref{eq:36})--(\ref{eq:2.3})
using finite element method. First, we need to get a variational form
of problem (\ref{eq:36})--(\ref{eq:2.3}). Let us introduce standard
Hilbert space of scalar functions $L_2(\Omega)$ with the following
inner product and norm:
\[
(u, v) = \int_\Omega u(\vec{x}) v(\vec{x}) d\vec{x}, \quad \| u \| =
(u, u)^{1/2}.
\]
For vector functions we define $\mathbf{L}_2(\Omega) =
[L_2(\Omega)]^3$. In addition, let $H^1(\Omega)$ and
$\mathbf{H}^1(\Omega)$ be Sobolev spaces of scalar and vector
functions, respectively.

Further, we define the spaces of trial and test functions as follows:
\begin{align*}
  \mathbf{V} &= \{ \vec{v} \in \mathbf{H}^1(\Omega)\ : \
  \vec{v}(\vec{x}) = 0, \ \vec{x} \in \Gamma_2 \} .
\end{align*}
Multiplying both sides of (\ref{eq:36}) by $\vec{v} \in \mathbf{V}$
and integrating the result by part, we get:
\begin{equation}
  \label{eq:4.1}
  \int_{\Omega} \tensor{\sigma}(\vec{u}) \tensor{\varepsilon}(\vec{v})
  d\vec{x} = \int_{\Gamma_3} (\vec{f}, \vec{v}) d\vec{s},
\end{equation}
where $\vec{u} \in \mathbf{V}$.

For numerical solving we need to transfer continuous variational
problem(\ref{eq:4.1}) to discrete one. Let tetrahedron mesh $\Omega_h$
is generated.  Further on the mesh $\Omega_h$ we introduce
\cite{Hughes2012} finite-dimensional subspaces of trial and test
functions: $\mathbf{V}_h \subset \mathbf{V}$ and define the following
discrete variational problem: find $\vec{u}_h \in \mathbf{V}_h$ such
that
\begin{equation}
  \label{eq:4.2}
  \int_{\Omega} \tensor{\sigma}(\vec{u}_h) \tensor{\varepsilon}(\vec{v}_h)
  d\vec{x} = \int_{\Gamma_3} (\vec{f}, \vec{v}_h) d\vec{s},
  \quad  \forall \vec{v}_h \in \mathbf{V}_h .
\end{equation}
Solver of the obtained discrete problem (\ref{eq:4.2}) is implemented
using library \textsf{DOLFIN} \cite{LoggWellsEtAl2012a} of the
software package \textsf{FEniCS}\footnote{\url{http://fenicsproject.org}}.

\section{Software description}

We develop software \textsf{Computer Dental System} (\textsf{CDS})
designed for use by prosthodontists at various levels of dental
treatment.

\subsection{Main elements} 

The presented software package allows to carry out individual
biomechanical analysis of dental system of patient on the basis of CT
images. Originality of the software lies in the fact that it combines
the following functionality: medical records, imaging with
construction of the digital models and numerical simulation of
stress-strain state of dental system based on which treatment
(prosthetics) option is chosen.

Graphical user interface is implemented using
\textsf{Qt}\footnote{\url{http://qt.io}}, crossplatform framework for
development software in \textsf{C++}
\cite{book:246597,summerfield2011advanced}. In addition for
visualization and processing 3D objects we use \textsf{Visualization Toolkit}
\textsf{(VTK})\footnote{\url{http://www.vtk.org}}, crossplatform
applied software for 3D modelling and visualization \cite{VTKbook2006}.

The main window of the presented software contains the following tabs:
\begin{itemize}
\item \emph{Patients}
\item \emph{Segmentation}
\item \emph{Prostheses}
\item \emph{Calculation}
\end{itemize}

\begin{figure}[h]
  \centering
  \includegraphics[width=1.0\textwidth]{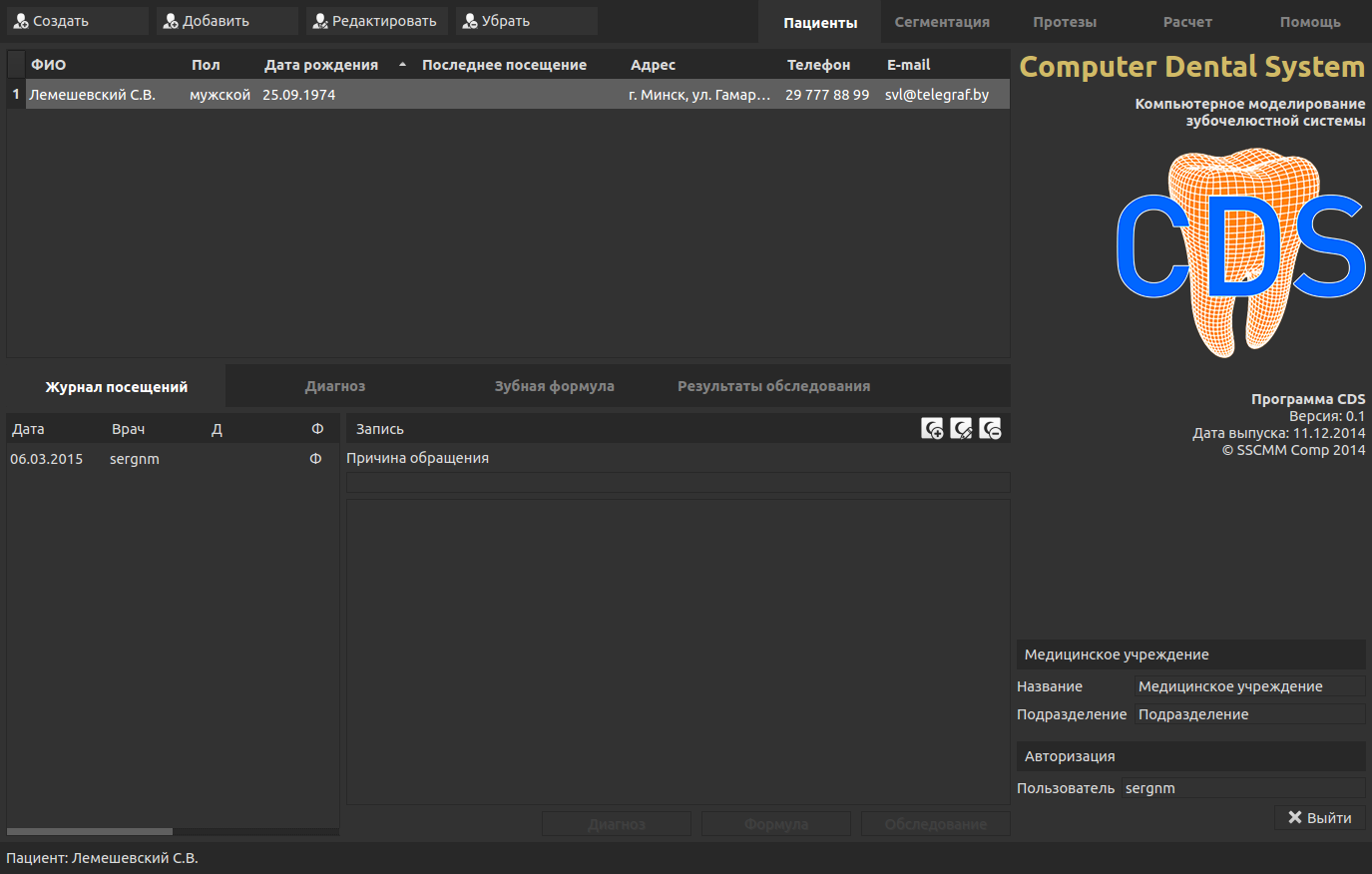}
  \caption{Tab  \emph{Patients}}
  \label{fig:3}
\end{figure}

First, a doctor performs inspection of patient: carries out external
examination, determines the state of the temporomandibular joint,
estimates the occlusion, the oral mucosa and the state of periodontal ligament
and hard tissue of teeth. The results of the examination are recorded
in the patient's history, which is filled on the tab  \emph{Patients}
(see Figure~\ref{fig:3})

According to the results of examination doctor makes the diagnosis and
preliminary treatment plan. At this stage, it is necessary to solve
the questions concerning the conduct of professional oral hygiene,
extraction of teeth that cannot be treated and restored. In addition,
if necessary, actions on surgical preparation of the oral cavity for
prosthetics are determined. Plan of endodontic treatment of the
supporting teeth is drown up only after final selection of the design of
prosthesis. At this stage doctor also preselect several future designs
of the prostheses.

\subsection{Computational kernel} 

\begin{figure}[h]
  \centering
  \includegraphics[width=1.0\textwidth]{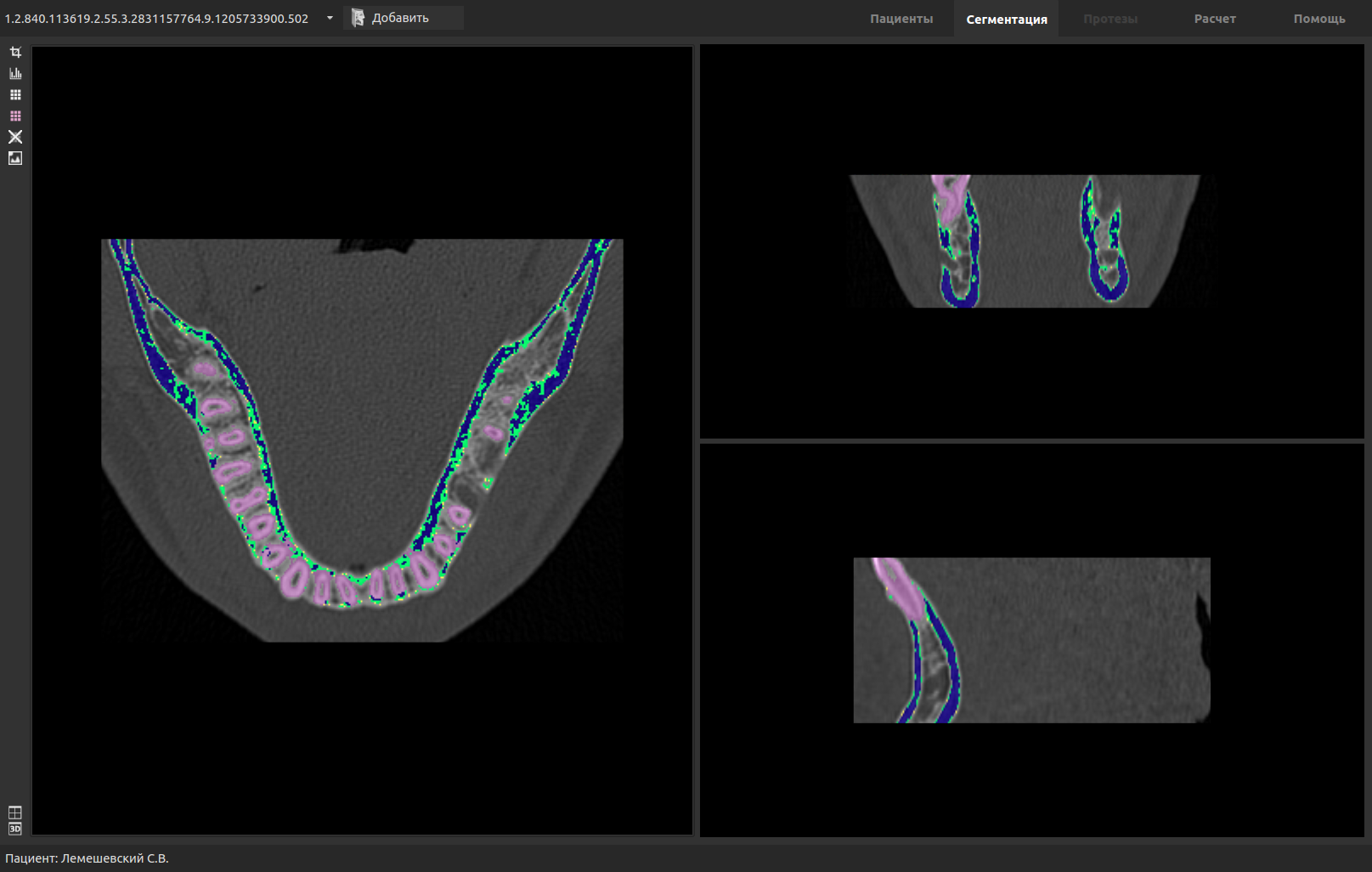}
  \caption{Tab \emph{Segmentation}}
  \label{fig:4}
\end{figure}

Biomechanical analysis and further choice of the design of prosthesis
is performed using 3D digital model of the dental system of
patient. Digital model constructed as described above includes all
teeth and jawbone. Figure~\ref{fig:4} illustrates the tab
\emph{Segmentation}, where doctor can construct the digital geometric
model using CT images.

Because of insufficient resolution of contemporary CT scanners
periodontal ligament is not contrasted in the image. Therefore, domain of
periodontal ligament constructed when mesh is generated after doctor specify
the possible prosthesics.

\begin{figure}[h]
  \centering
  \includegraphics[width=1.0\textwidth]{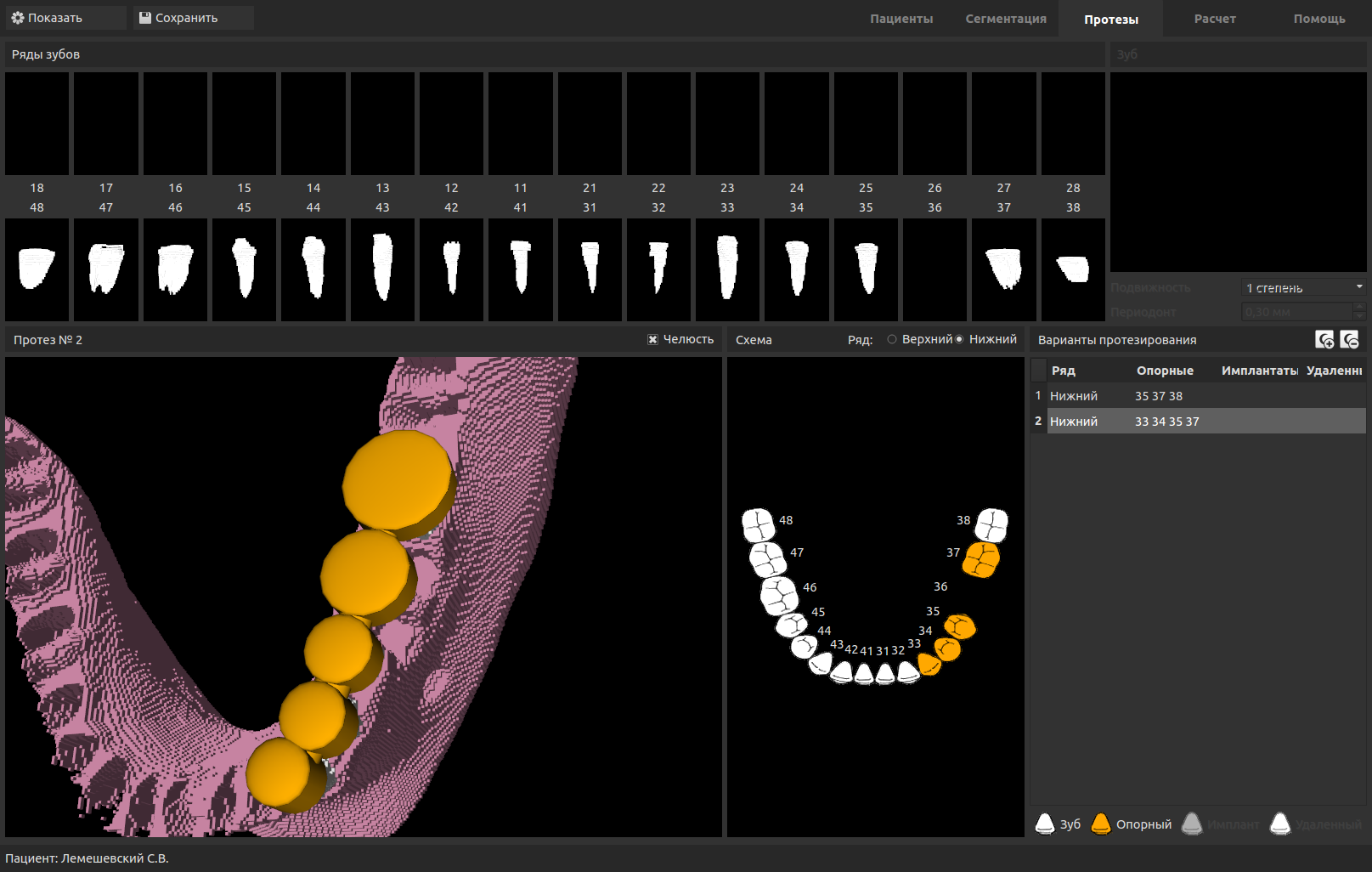}
  \caption{Tab \emph{Prostheses}}
  \label{fig:5}
\end{figure}

When the geometric models of jawbone and teeth are constructed, doctor
proceeds to construction of prostheses that have been identified at
the stage of examination and diagnosis. This feature is implemented on
the tab \emph{Prostheses} shown in Figure~\ref{fig:5}. Here doctor can add
prostheses selecting supporting teeth. Moreover, for each tooth doctor can
set the degree of mobility and thickness of periodontal ligament. Each degree
of mobility corresponds to its own Lam\`e coefficients in the domain
of periodontal ligament adjacent to corresponding tooth.

\begin{figure}[h]
  \centering
  \includegraphics[width=1.0\textwidth]{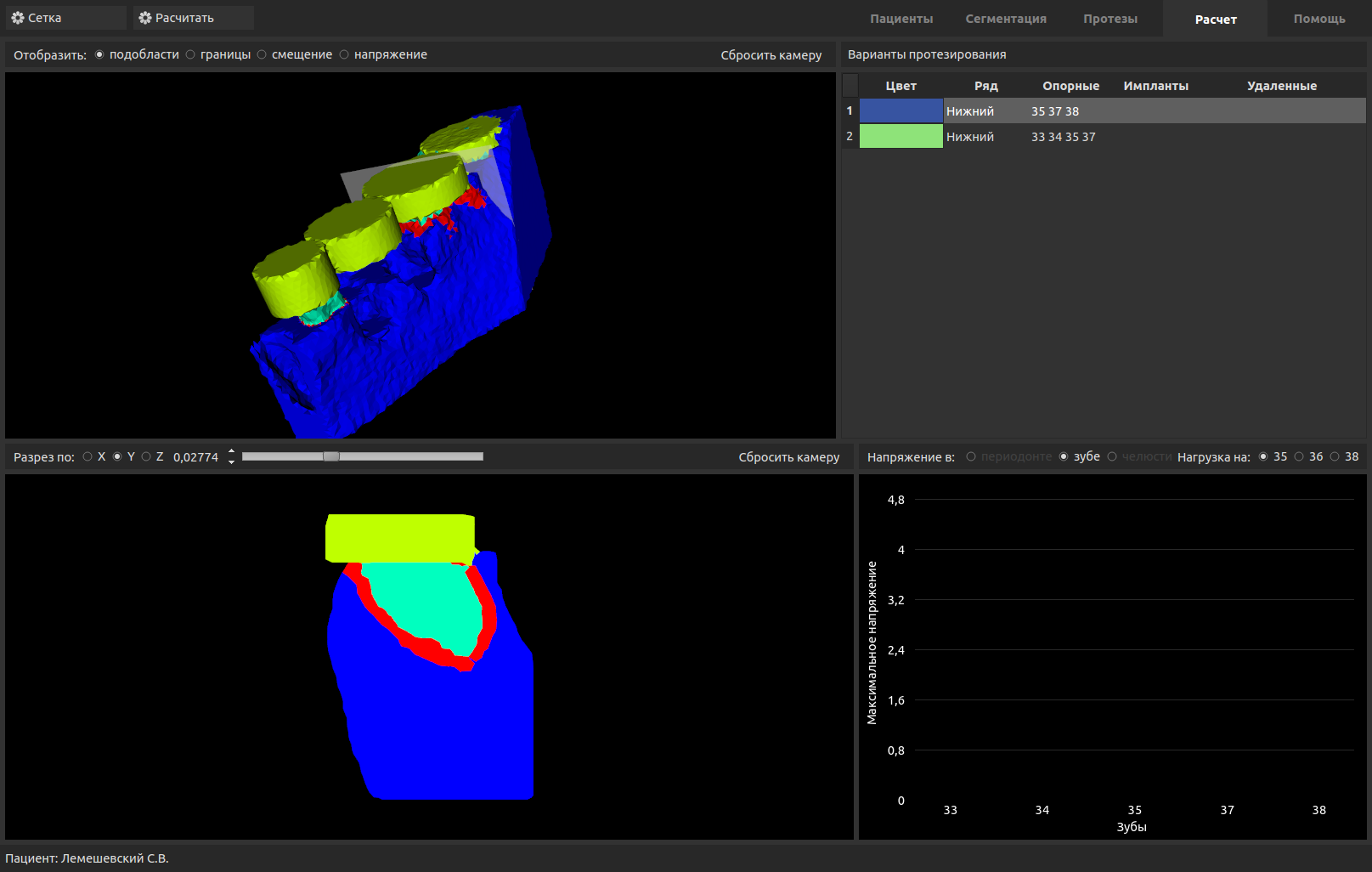}
  \caption{Tab \emph{Calculation}}
  \label{fig:6}
\end{figure}

When possible prostheses are constructed doctor can go to
performing calculation and analysis of the results. This actions are
performed on the tab \emph{Calculation} (Figure~\ref{fig:6}).

When we select tab \emph{Calculation}, it is automatically generated
using constructed prostheses the modelling domains: only part of
jawbone and supporting teeth are cut out and prostheses are added.

\begin{figure}[h]
  \centering
  \includegraphics[width=1.0\textwidth]{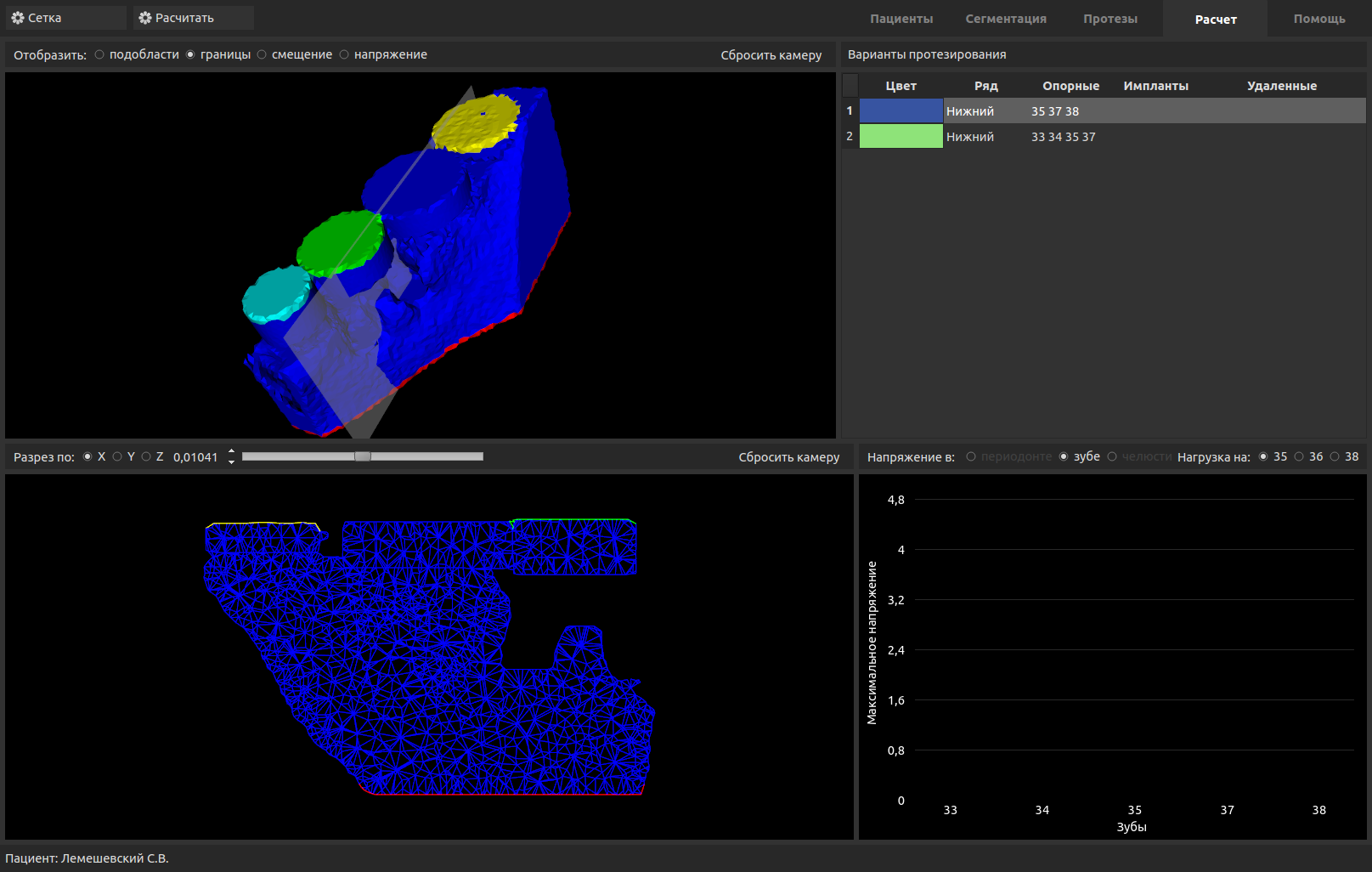}
  \caption{Boundary surfaces where loads are given}
  \label{fig:7}
\end{figure}

Here doctor generate computational meshes for all options of
prosthetics. In addition, domain of periodontal ligament with the given
thickness is also generated. Moreover, the standard surfaces of loads
are automatically marked: outer supporting teeth and center of prosthesis
(see Figure~\ref{fig:7}). The standard load is equal to 100 MPa and
normal to the surface. Such loads are equivalent to force of 10 kg
required for mastication.

On the tab \emph{Calculation} the results are visualized for performing
biomechanical analysis. After biomechanical analysis of various
designs of prostheses and comparing the obtained stress-strain states,
doctor choose the design of prosthesis with the less negative impact
on dental system. For biomechanical analysis the fields of strains and
stresses are visualized. In addition, it is implemented displaying
slices along planes. Moreover, the maximum near the teeth of the
computed fields are shown.

\begin{figure}[h]
  \centering
  \includegraphics[width=1.0\textwidth]{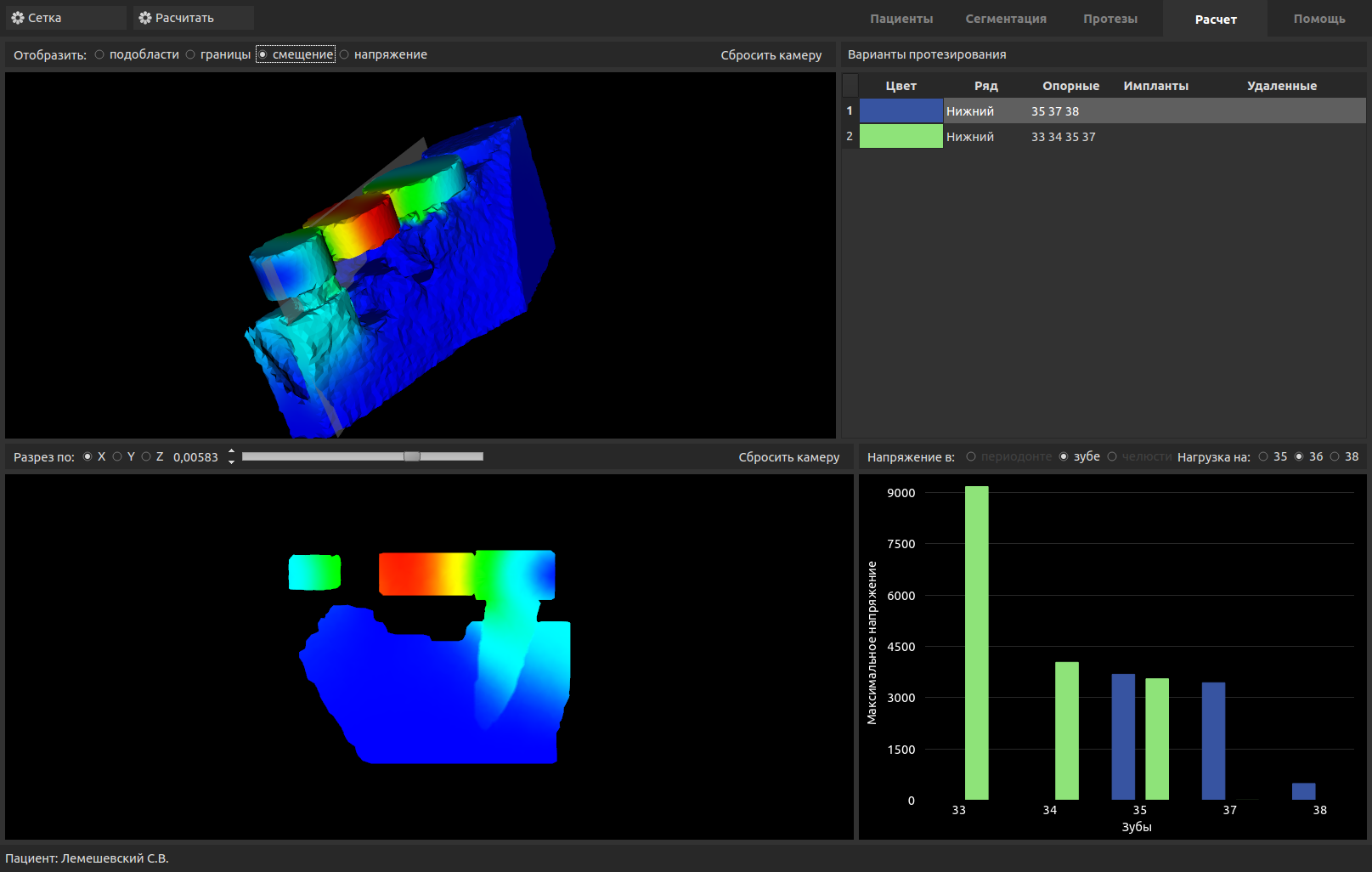}
  \caption{Calculation results: strain field}
  \label{fig:8}
\end{figure}

\begin{figure}[h]
  \centering
  \includegraphics[width=1.0\textwidth]{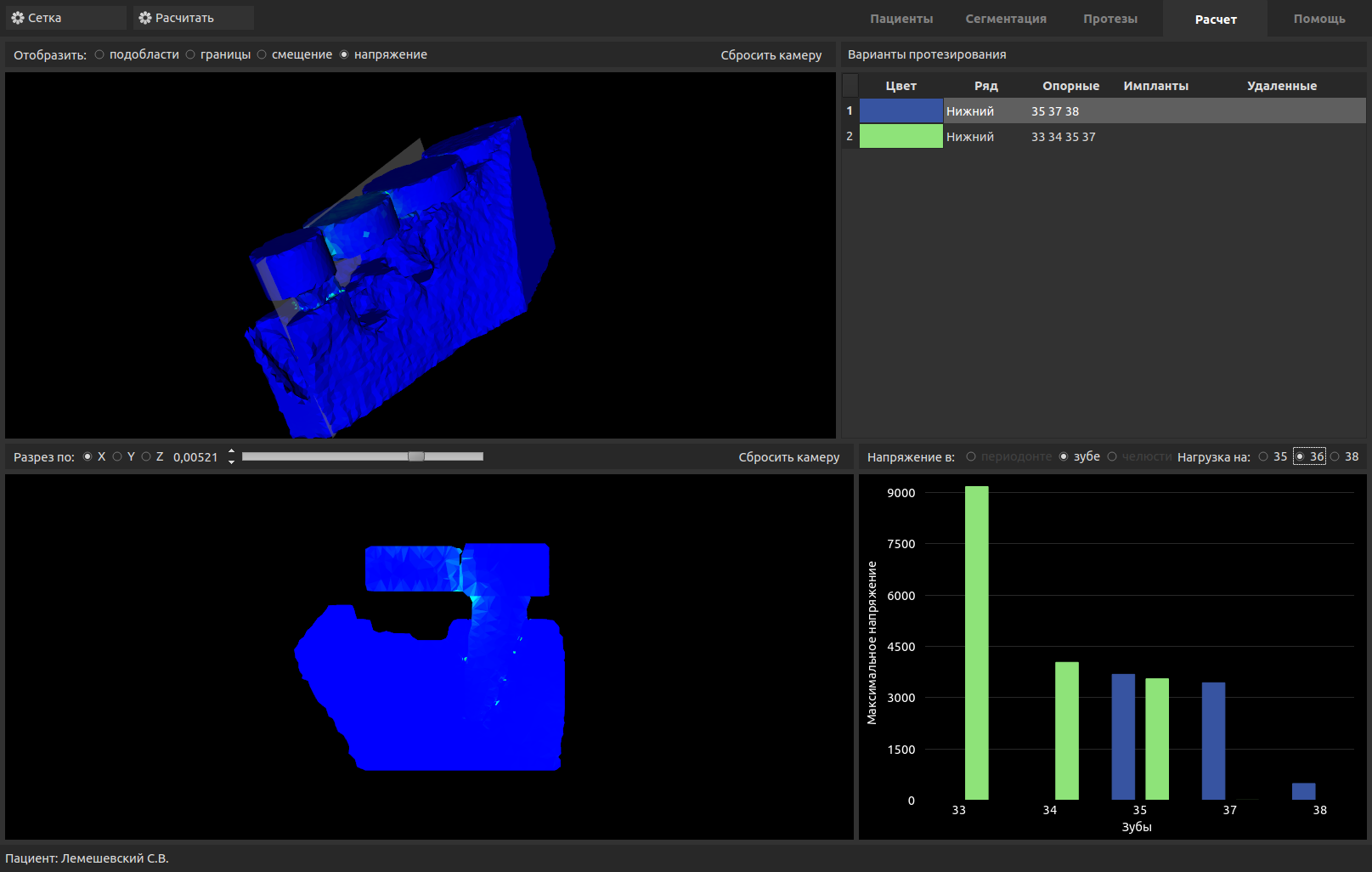}
  \caption{Calculation results: stress field}
  \label{fig:9}
\end{figure}

Analysis of the proposed design of prosthesis is carried out on the
basis of investigation of the strains and stresses (see
Figures~\ref{fig:8} and \ref{fig:9}). 

\clearpage

\section*{Acknowledgements}

This work was supported by the Russian Foundation for Basic Research
(project 14-01-00785), the National Academy of Sciences of Belarus
(project \emph{Convergence 1.5.01}), the SCST of the Republic of Belarus (project 20092519).

\bibliography{Vab}
\end{document}